\def\rfr#1{eq. (\ref{#1})}

\def\derp#1#2{\rp{\partial{#1}}{\partial{#2}}}
\def\dert#1#2{\frac{{{d}}{#1}}{{{d}}{#2}}}              

\def\asec{$''$ cy$^{-1}$}

\def\bar{\begin{eqnarray}}
\def\ear{\end{eqnarray}}
\def\bb{\bibitem}
\def\eqi{\begin{equation}}
\def\eqf{\end{equation}}
\def\eqia{\begin{eqnarray}}
\def\eqfa{\end{eqnarray}}
\def\rp#1#2{{#1\over#2}}
\def\ct#1{\cite{#1}}
\def\lb#1{\label{#1}}

\def\asec{$''$ cy$^{-1}$}

\def\oc2{$\mathcal{O}(c^{-2})$}

\def\spn{\boldsymbol{S}}

\documentclass[11pt]{article}
\usepackage{amsmath,amsthm,amscd,amssymb}
\usepackage{latexsym}
\usepackage{graphicx,epsfig}

\begin{document}

\noindent{\bf \LARGE{Spin precession in the
Dvali-Gabadadze-Porrati braneworld scenario }}
\\
\\
\\
{Lorenzo Iorio, {\it FRAS, DDG}}\\
{\it
\\Viale Unit$\grave{a}$ di Italia 68, 70125\\Bari, Italy
\\e-mail: lorenzo.iorio@libero.it}

\begin{abstract}
In this letter we work out the secular precession of the spin of a
gyroscope in geodesic motion around  a central mass in the
framework of the Dvali-Gabadadze-Porrati multidimensional gravity
model. Such an effect, which depends on the mass of the central
body and on the orbit radius of the gyroscope, contrary to the
precessions of the orbital elements of the orbit of a test body,
is far too small to be detected.
\end{abstract}
\section{Introduction}
The Dvali-Gabadadze-Porrati (DGP) multidimensional gravity model
\cite{DGP00} has recently attracted great interest because it not
only allows to explain the observed acceleration of the expansion
of our Universe but also predicts some tiny post-Einsteinian
effects that are testable at local scales and yield information on
the global properties of the Universe as the kind of cosmological
expansion currently ongoing \ct{DGZ03}. For a comprehensive
phenomenological overview of the DGP gravity see \ct{Lue05}.

Up to now, secular precessions of the longitude of pericentre
$\varpi$ \ct{LS03} and of the mean anomaly $\mathcal{M}$
\ct{Iorio05a} of the orbit of a test particle  freely falling
around a central body have been worked out. Such effects, which
depend on the eccentricity $e$ via second-order terms and are
independent of the semimajor axis $a$ of the orbiter, amount to
$10^{-4}-10^{-3}$ arcseconds per century (\asec). The ideal
test-bed for them is represented by the inner planets of the Solar
System \ct{Iorio05b, Iorio05c}. Such orbital precessions lie at
the edge of the present-day precision of the latest planetary
data. The DGP features of motion related to the self-acceleration
cosmological phase are compatible to them \ct{Iorio05d}.

In this letter we wish to investigate the impact of DGP gravity on
the precession of a spin $\spn$ orbiting a central body of mass
$M$.
\section{The spin precession}
In the DGP picture our Universe is a (3+1) space-time brane
embedded in a larger five-dimensional Minkowskian bulk. Contrary
to the other forces constrained to remain on the brane, gravity
can fully explore the entire bulk getting strongly modified at
scales $r\geq r_c$, where $r_c$ is a free-parameter fixed by the
observations of the Supernov${\rm \ae}$ Type IA to $r_c\sim 5$
Gpc. At much smaller scales $r\ll r_c$ the usual Newton-Einstein
gravity is recovered apart from small corrections. By neglecting
the effects of the spatial curvature, the modifications to the
Newtonian potential can be accounted for as \ct{LS03} \eqi
(ds)^2=\left(1-\rp{R_g}{2r}\pm\sqrt{\rp{R_g r
}{2r^2_c}}\right)^2(cdt)^2-(dx)^2-(dy)^2-(dz)^2,\lb{metrica}\eqf
where $R_g=2GM/c^2$ is the Schwarzschild radius of the central
mass and the $\pm$ sign is related to the different cosmological
phases: the $+$ sign is for the Friedmann-Lema${\rm
\hat\i}$tre-Robertson-Walker (FLRW) phase while the $-$ sign is
for the self-accelerated phase.

Let us consider a gyroscope falling along a geodesic of the
space-time metric: thus, its spin vector is carried along by
parallel transport according to \ct{OhRu94}
\eqi\dert{S^{\mu}}{\tau}=-\Gamma^{\mu}_{\alpha\beta}S^{\alpha}\dert{x^{\beta}}{\tau},\eqf
where $\Gamma^{\mu}_{\alpha\beta}$ are the Christoffel symbols
\eqi\Gamma^{\mu}_{\alpha\beta}=\rp{g^{\mu\nu}}{2}\left(\derp{g_{\nu\alpha}}{x^{\beta}}+\derp{g_{\beta\nu}}{x^{\alpha}}
-\derp{g_{\alpha\beta}}{x^{\nu}}\right)\eqf and $\tau$ is the
proper time along the geodesic of the orbiting gyroscope.

By neglecting all the terms proportional to $R^2_g,\ R_g/r^2_c,\
R_g^{3/2}/r_c$ we can pose
\begin{eqnarray}
g_{00}&\simeq & 1-\frac{R_g}{r}\pm\sqrt{\frac{2 R_g r}{r_c^2}},\\
g^{00}&\simeq & 1+\frac{R_g}{r}\mp\sqrt{\frac{2 R_g r}{r_c^2}}.
\end{eqnarray}
The non-vanishing Christoffel symbols for the metric of
\rfr{metrica} are \eqi \Gamma^{0}_{0i}=\Gamma^{i}_{00}\simeq
\frac{R_g}{2}\frac{
x^{i}}{r^3}\pm\sqrt{\frac{R_g}{8r_c^2}}\frac{x^{i}}{r^{3/2}},\
i=1,2,3. \eqf The equations for the spatial components of the spin
thus become \eqi
\dert{S^i}{\tau}=-\Gamma^{i}_{\alpha\beta}S^{\alpha}\dert{x^{\beta}}{\tau}=-\Gamma^{i}_{00}S^{0}c,\
i=1,2,3. \lb{vriz}\eqf The time-like component $S^0$ of the spin
four-vector is, from $g_{\mu\nu}S^{\mu}\dert{x^{\nu}}{s}=0$ \eqi
S^0=-\frac{1}{g_{00}}\left(S_x g_{11}\dert{x}{s}+S_y
g_{22}\dert{y}{s}+S_z
g_{33}\dert{z}{s}\right)\simeq\left(1+\frac{R_g}{r}\mp\sqrt{\rp{2R_g
r }{r_c^2}}\right)\rp{\spn\cdot\boldsymbol v}{c}.\eqf The spin
precession due to the DGP correction to the Newtonian potential is
thus \eqi\dot\spn_{\rm DGP}=\mp\sqrt{\frac{R_g}{8r_c^2
r^3}}(\spn\cdot\boldsymbol v)\boldsymbol r.\lb{formulaz}\eqf

By considering a circular planar orbit of radius $a$ and period
$P=2\pi\sqrt{a^3/GM}\doteq 2\pi/n$, so that
\begin{eqnarray}
\boldsymbol r&=& a\cos nt\boldsymbol i+a\sin nt\boldsymbol j,\\
\boldsymbol v&=& -v\sin nt\boldsymbol i+v\cos nt \boldsymbol j,
\end{eqnarray}
and averaging over one orbital period \rfr{formulaz} becomes \eqi
\langle\dot\spn_{\rm DGP}\rangle=\pm\sqrt{\frac{R_g}{32r_c^2
a}}v\left(-S_y\boldsymbol i+S_x\boldsymbol
j\right)=\pm\sqrt{\frac{R_g}{32r_c^2 a^3}}(\boldsymbol
r\times\boldsymbol v )\times\spn\doteq\boldsymbol\Omega_{\rm DGP
}\times\spn.\eqf Thus, by assuming $v=na$, the spin of an orbiting
gyroscope moving along a circular orbits precesses at a speed
\eqi\Omega_{\rm DGP}=\rp{GM}{4c r_c a}.\lb{prcs}\eqf Contrary to
the secular precessions of the Keplerian orbital elements of a
test particle, which is determined from the geodesic equation of
motion, such a spin precession depends on the mass of the central
body and on the radius of the gyroscope's orbit. Unfortunately,
the size of such an effect amounts to only $10^{-11}-10^{-12}$
\asec\ for the orbital angular momenta of the inner planets of the
Solar System and to $10^{-12}$ milliarcseconds per year for the
gyroscopes of the GP-B spacecraft \cite{Eve01}. It is so because
of the $c^{-1}$ factor in \rfr{prcs} due to the presence of
$dx^0/d\tau$ only once in \rfr{vriz}. Instead, in  the geodesic
equation of motion $dx^0/d\tau$ appears twice.

\section{Conclusions}
In this letter we have worked out the behavior of the spin of a
gyroscope in geodesic motion around a central mass in the
framework of the Dvali-Gabadadze-Porrati braneworld scenario. It
turns out that the spin undergoes a precession which depends both
on the mass of the central body and on the radius of the gyroscope
orbit, contrary to the secular precessions of the Keplerian
orbital elements of a test particle. The magnitude of such an
effect is too small to be detected in a foreseeable future.

\end{document}